\documentclass[11pt]{article}

\usepackage[top=1in, bottom=1in, left=1in, right=1in]{geometry}

\usepackage{amsmath}
\usepackage{amsfonts}
\usepackage{wasysym}
\usepackage{graphicx}
\usepackage{comment}

\usepackage[compress,square,numbers]{natbib}

\usepackage{hyperref}

\def\vect{\vec}

\def\filetype{eps}

\begin{document}


\thispagestyle{empty}

$\,$

\addvspace{50pt}

\begin{center}

\Large{\textbf{Cosmological Inhomogeneities with Bose-Einstein Condensate Dark Matter}}
\\[35pt]
\large{Ben Kain$^1$ and Hong Y. Ling$^2$}
\\[20pt]
\textit{$^1$Department of Physics}
\\ \textit{College of the Holy Cross}
\\ \textit{Worcester, MA 01610, USA}
\\[10pt]
 \textit{$^2$Department of Physics and Astronomy}
\\ \textit{Rowan University}
\\ \textit{Glassboro, NJ 08028, USA}
\end{center}

\addvspace{35pt}

\begin{abstract}
\noindent We consider the growth of cosmological perturbations to the energy density of dark matter during matter domination when dark matter is a scalar field that has undergone Bose-Einstein condensation.  We study these inhomogeneities within the framework of both Newtonian gravity, where the calculation and results are more transparent, and General Relativity.  The direction we take is to derive analytical expressions, which can be obtained in the small pressure limit.  Throughout we compare our results to those of the standard cosmology, where dark matter is assumed pressureless, using our analytical expressions to showcase precise differences.  We find, compared to the standard cosmology, that Bose-Einstein condensate dark matter leads to a scale factor, gravitational potential and density contrast that increase at faster rates. 
\end{abstract} 

\newpage


\section{Introduction}

Astrophysical observations indicate that 23\% of the energy density of the Universe is of an unknown nonbaryonic form, known as dark matter.  While a precise explanation remains elusive, a weakly interacting, nonrelativistic massive particle is favored.  Leading candidates include supersymmetric particles, such as neutralinos, and the axion, originally proposed to solve the strong CP problem in QCD.

Dark matter is often modeled as a pressureless, nonrelativistic particle, known as cold dark matter.  While this model has achieved significant success, in particular with early universe and large-scale cosmology, it meets with difficulty on galactic scales.  Cold dark matter simulations of galactic halo formation predict density profiles with a central cusp \cite{navarro}, while observations indicate constant density cores \cite{burkert}.  Scalar field dark matter that has undergone Bose-Einstein condensation \cite{sin,leekoh,hubak,silmal,wang,mfs,sikyan,sahni} has been considered as a solution to this problem since the resulting density profiles agree with observed rotation curves \cite{hubak,BH,harkocusp, chavanis2}.

At low temperatures, where wave aspects dominate, a many-body system of bosons exhibits Bose enhancement, whereby bosons favor joining highly populated states.  As a result, bosons can pile up in the same ground state forming a coherent matter wave of macroscopic size known as a Bose-Einstein condensate (BEC).  This will occur when the thermal de Broglie wavelength, $\lambda = \sqrt{2\pi\hbar^2/2mk_B T}$, begins to exceed the interparticle spacing, $n^{-1/3}$, so that wave functions of individual bosons begin to overlap, where $m$ is the boson mass and $n$ its number density.  Equivalently, Bose-Einstein condensation occurs when the temperature, $T$, drops below the critical temperature,
\begin{equation}
	T_c = \frac{1}{m} \frac{2\pi \hbar^2}{k_B [\zeta(3/2)]^{2/3}} n^{2/3},
\end{equation}
where $\zeta(x)$ is the Riemann-Zeta function.  

The experimental realization of trapped BECs in dilute alkaline atoms in 1995 \cite{anderson} has led to a renewed interest in BECs, which has been a subject that unifies many disciplines, for example neutron stars \cite{glendenning}, superconductivity \cite{mahan} and, what is our interest here, dark matter \cite{sin,leekoh,hubak,silmal,wang,mfs,sikyan,sahni}.  A great virtue in the study of BECs is that in ultracold atomic systems important system parameters, including the interaction strength between particles and the dimensionality of the system, can be tuned precisely.  This opens up the possibility of using ultracold atomic systems in the laboratory to simulate phenomena on galactic and cosmological scales.  Possible examples include creating a system where the BEC is subject to so-called electromagnetically induced ``gravity" with a $1/r$ interatomic attractive potential \cite{odell} and creating a controlled explosion of atoms by suddenly making the s-wave scattering length negative, a phenomenon dubbed ``bosenova" because of its resemblance, on a vastly lower energy scale, to the core collapse in a supernova \cite{donley}.

Our focus here is with the cosmological applications of BECs, in particular BEC dark matter.  In addition to the density profile and rotation curves mentioned above, investigations of BEC dark matter include the study of vortex formation \cite{silmal, yumor, brook, kl, shapiro}, additional aspects of galactic structure \cite{galactic}, Bose-Einstein condensation in the early universe \cite{urena,fukuyama,harko2} and axions \cite{sikivie}.  Recently a study of its cosmology was initiated by Harko \cite{harko} and Chavanis \cite{chavanis} (see also \cite{fukuyama,velten,kamion,khlopov}). They investigated the evolution of inhomogeneities in the dark matter energy density.  Such inhomogeneities eventually become nonlinear and lead to galaxy formation and indirectly affect anisotropies in the cosmic microwave background radiation.  They derived evolution equations for the density contrast in Newtonian gravity \cite{chavanis} and post-Newtonian gravity \cite{harko,chavanis} and presented numerical solutions to these equations.

In this work we also analyze the evolution of cosmological inhomogeneities after dark matter has undergone Bose-Einstein condensation.  Our departure from previous results is twofold.  First, we present simple analytical solutions which allow for a precise understanding of how BEC dark matter differs from standard cold dark matter.  Since dark matter is believed to have a small pressure (indeed, it is often modeled as having zero pressure), by taking the small pressure limit such analytical solutions are obtainable.  Second, while we will begin with Newtonian gravity, we derive the evolution equations using the complete theory of General Relativity.  The use of General Relativity is necessary when considering superhorizon perturbations, which are beyond the reach of Newtonian gravity.  It is also necessary for considering anisotropies in the radiation spectrum.  We do not study anisotropies here, as it is outside the scope of our work, but such a study would be interesting and important and would rely on our results.

In the next section we review the Gross-Pitaevskii equation coupled to the Poisson equation and the Thomas-Fermi approximation, which has become the standard framework for describing a gravitating BEC.  We then derive the equation of state for BEC dark matter and use it to analyze the homogeneous, unperturbed cosmology.  In section \ref{inhomogeneities} we derive the evolution equations for inhomogeneities during matter domination, first in Newtonian gravity and then in General Relativity.  Throughout we compare our results to those of the standard cosmology, where dark matter is pressureless.  When making these comparisons we shall refer to such dark matter as standard cold dark matter (SCDM).  We conclude in section \ref{conclusion}.


\section{Bose-Einstein Condensate Dark Matter}
\label{sec:BECDM}


\subsection{Hydrodynamic Description}
\label{sec:hydro}

We assume that dark matter is composed of scalar bosons, of mass $m$, that have undergone a phase transition toward Bose-Einstein condensation.  To describe the BEC we employ the standard symmetry-breaking mean field approach, which is expected to be valid for systems with a sufficiently large number of particles and at temperatures far below the BEC transition temperature.  In this approach we may start with the Gross-Pitaevskii energy functional:
\begin{equation}\label{BECaction}
	E[\psi]=\int d^3 x \left[ \frac{\hbar^2}{2m} |\nabla \psi(t,\vec{x})|^2 + \frac{1}{2} V_0 |\psi(t,\vec{x})|^4 + \frac{1}{2}m V_G(t,\vec{x}) |\psi(t,\vec{x})|^2 \right],
\end{equation}
where $\psi(t,\vect{x})$ is the order parameter, or macroscopic wave function, describing the BEC and is normalized such that $|\psi|^2$ is the number density.  The first term is the standard kinetic energy term of nonrelativistic quantum mechanics.  The second term represents a quartic, contact interaction with strength
\begin{equation} \label{V0}
	V_0 = \frac{4\pi \hbar^2 a_s}{m},
\end{equation}
where $a_s$ is the s-wave scattering length, which we take to be positive ($a_s > 0$).  The third term is the gravitational potential,
\begin{equation}
	V_G(t,\vect{x}) = -G m \int d^3 x' \frac{|\psi(t,\vect{x}')|^2} {|\vect{x}-\vect{x}'|^2},
\end{equation}
which satisfies Poisson's equation:
\begin{equation} \label{Poisson}
	\nabla^2 V_G(t,\vect{x}) = 4\pi Gm|\psi(t,\vect{x})|^2.
\end{equation}
By including the gravitational potential, (\ref{BECaction}) describes a BEC coupled to gravity. 

In the mean field approach one ignores high order correlations due to bosonic quantum field fluctuations.  This allows the BEC to be described by the equations of motion that follow from variation of (\ref{BECaction}), under the constraint that the total number of particles is conserved:
\begin{equation}\label{GP}
	i\hbar \frac{\partial \psi}{\partial t} = -\frac{\hbar^2}{2m}\nabla^2 \psi + V_0 |\psi|^2 \psi + m V_G\psi - \mu \psi,
\end{equation}
where $\mu$ is the chemical potential.  This equation is known as the Gross-Pitaevskii equation.  It may be written in the hydrodynamic representation, which is more useful for our purposes, by separating the wave function into its modulus and phase,
\begin{equation}
	\psi(t,\vect{x}) = |\psi(t,\vect{x})| e^{iS(t,\vect{x})},
\end{equation}
both of which are real, and then describing the condensate in terms of its energy density and local velocity:
\begin{equation} \label{rhoveldef}
	\rho(t,\vect{x}) = mc^2|\psi(t,\vect{x})|^2, \qquad v(t,\vect{x}) = \frac{\hbar}{m}\nabla S(t,\vect{x}).
\end{equation}
In terms of these variables the Gross-Pitaevskii equation (\ref{GP}) and the Poisson equation (\ref{Poisson}) become
\begin{subequations}\label{eqs}
\begin{align}
	\frac{\partial\rho}{\partial t} &= -\nabla \cdot \left(\rho \vect{v} \right) \label{eqcontin}\\
	-\frac{\partial \vect{v}}{\partial t} &= -\frac{\hbar^2}{2m^2}\nabla \left(\frac{1}{\rho} \nabla^2\rho \right) + \frac{1}{2}\nabla (\vect{v}^2) + \frac{V_0}{m^2 c^2}\nabla \rho + \nabla V_G \label{eqsb}\\
	\nabla^2 V_G &= \frac{4\pi G}{c^2}\rho. \label{poiseq}
\end{align}
\end{subequations}
Aside from the first term on the right hand side of (\ref{eqsb}), the top two equations comprise the hydrodynamic description of the condensate since (\ref{eqcontin}) is the continuity equation and (\ref{eqsb}) is the Euler equation from classical fluid dynamics.    The first term on the right hand side of (\ref{eqsb}) is called the quantum pressure term.  Unfortunately this term often makes analytic solutions difficult to come by.  It may be traced to the first term in (\ref{BECaction}), which originates from the uncertainty principal and hence cannot find its analog in classical physics.  Because this term contains the gradient of the energy density, as the number of particles in the condensate, or equivalently the size of the wave function, increases, the quantum kinetic energy becomes negligible compared to other energy contributions except near boundaries of the condensate.  Neglecting the quantum pressure term is known as the Thomas-Fermi approximation (for a more precise definition of the gravitational Thomas-Fermi regime see \cite{kl}).  It is a common practice to employ the Thomas-Fermi approximation in the study of perturbations to BEC densities \cite{dalfovo}, something we shall do in the next section.


\subsection{Equation of State}
\label{sec:eos}

The equation of state for a fluid relates the pressure to the energy density, $p=p(\rho)$, and is of fundamental importance in cosmology.  It may be obtained for a homogeneous BEC by ignoring the gravitational potential in (\ref{BECaction}) and taking $\psi(t,\vect{r})=\psi_0$ to be real and constant, where $\psi_0^2 = N/V$ is the number density.  From (\ref{BECaction}) the BEC has energy
\begin{equation}
	E = V \left( \frac{1}{2} V_0 \psi_0^2 \right) = \frac{1}{2} V_0 \frac{N^2}{V}.
\end{equation}
The pressure is then
\begin{equation}
	p = -\left.\frac{\partial E}{\partial V} \right|_{N} = \frac{1}{2} V_0 \psi_0^4.
\end{equation}
In a BEC each particle contributes an energy $mc^2$, so the energy density is $mc^2$ times the number density, $\rho=mc^2\psi_0^2$, and we find the equation of state
\begin{equation} \label{BECeos}
	p = \frac{V_0}{2m^2 c^4} \rho^2 = \frac{2\pi\hbar^2 a_s}{m^3 c^4}\rho^2 \equiv \lambda \rho^2,
\end{equation}
where $\lambda = 2\pi\hbar^2 a_s /m^3 c^4$.

Dark matter is thought to be cold and nearly pressureless.  The standard assumption is that it is a pressureless, perfect fluid, which we will refer to as standard cold dark matter (SCDM), with equation of state $p_\text{SCDM}=0$.  BEC dark matter has nonzero pressure and the nontrivial equation of state (\ref{BECeos}).  We will analyze BEC dark matter using the dimensionless quantity
\begin{equation} \label{eos}
	w \equiv \frac{p}{\rho} = \lambda\rho.
\end{equation}
When solving for approximate, analytical solutions, we will take $w$ to be a small perturbation around the SCDM solution $w_{\text{SCDM}}=0$.  The SCDM results can be obtained by setting $w=0$.

In general $w$ is not constant, but we will make use of it evaluated at its (constant) present-day value $w_0$.  Further, we can introduce the dark matter fraction $\Omega_{\text{DM}} = \rho_/\rho_{c}$, where $\rho_{c}$ is the critical energy density necessary for a flat universe \cite{kolb}.  Then
\begin{equation}
	w_0 = \lambda \Omega_{\text{DM},0} \rho_{c,0}
\end{equation}
where $ \Omega_{\text{DM},0}$ and $\rho_{c,0}$ are the present-day values.


\subsection{Homogeneous, Isotropic Cosmology}
\label{sec:homogcos}

If we ignore perturbations, the Universe is well known to be flat, isotropic and homogeneous on the distance scales of interest \cite{muk}.  It may be described by the evolution of the scale factor, $a(t)$, which evolves according to the Friedmann equations \cite{kolb},
\begin{equation} \label{Friedmann}
	H^2 = \frac{8\pi G}{3c^2}\rho_0, \qquad \dot{\rho_0} = -3 H\left(\rho_0 + p_0 \right),
\end{equation}
where $H=\dot{a}/a$ is the Hubble parameter and a dot denotes a time derivative.  Here, and from now on, the subscripted 0 on the energy density, $\rho_0$, and the pressure, $p_0$, indicate that these are unperturbed, background quantities.  The cosmology of BEC dark matter follows from the Friedmann equations and the equation of state.  In section \ref{inhomgNewton}, when studying inhomogeneities in Newtonian gravity, we will derive the Friedmann equations directly from (\ref{eqs}).  Here we simply quote their well known form.  

Our goal in this subsection is to determine the evolution of the scale factor during matter domination when BEC dark matter, with equation of state (\ref{eos}), dominates the total energy density of the Universe and perturbations have been ignored (perturbations will be considered in the next section).  Using (\ref{eos}) and the second equation in (\ref{Friedmann}) we have \cite{harko}
\begin{equation} \label{rhosol}
	\rho_0(a) = \frac{A}{a^3 - \lambda A},
\end{equation}
where $A$, since it is the exponential of an arbitrary constant, is positive, but otherwise arbitrary.  It may be fixed by requiring the energy density to have its present-day value, $\rho_0=\rho_{0,0}$, when the scale factor has its present-day value, $a=a_0$, leading to $A = \rho_{0,0} a_0^3/(1+\lambda \rho_{0,0})$.  With this we may rewrite (\ref{rhosol}) as \cite{harko,chavanis}
\begin{equation} \label{rhosolW}
	\rho_0(a) = \frac{\rho_{0,0}(1-W_0)}{(a/a_0)^3 - W_0},
\end{equation}
where for convenience we defined	$W_0 \equiv w_0/(1+w_0)$.  Using the solution (\ref{rhosolW}) and the first equation in (\ref{Friedmann}) we obtain \cite{harko}
\begin{equation} \label{ateosexact}
	\sqrt{\Omega_{\text{DM},0} (1-W_0)}H_0( t-t') = \frac{2\sqrt{W_0}}{3} \left(y - \tan^{-1} y\right),
\end{equation}
where $t'$ is an arbitrary constant, and
\begin{equation} \label{yeq}
	y \equiv \sqrt{\frac{1}{W_0} \left(\frac{a}{a_0}\right)^3 - 1}.
\end{equation}
$t'$ may be fixed by applying an initial condition.  Initially, in the very early Universe when the temperature was sufficiently high, BEC dark matter will not yet have condensed.  As the Universe expands and cools, dark matter will eventually begin to condense, taking a finite period of time to complete \cite{harko2}.  Once complete, (\ref{rhosolW}) is valid, so that completion occurs during $a>a_0 W_0^{1/3}$.  The initial condition we apply, which was used in \cite{harko,chavanis}, is $a(t=0)=a_0 W_0^{1/3}$, or equivalently $y(t=0)=0$, which sets $t'=0$.

So far everything we have done is exact.  We will now use the approximation $w_0 \ll 1$, which is the statement that the present-day pressure is small.  It follows from this that $W=w_0 + O(w_0^2)$ and $y \simeq \sqrt{(a/a_0)^3/w_0}$ is large.  Expanding around large $y$ and small $w_0$ we obtain
\begin{equation}
	\sqrt{\Omega_{\text{DM},0}} (H_0  t) = \frac{2}{3} \sqrt{w_0 +O(w_0^2)} \left[-\frac{\pi}{2} + y + \frac{1}{y} + O(y^{-3}) \right].
\end{equation}
This equation may be solved for $a$ and then expanded around small $w_0$ to obtain
\begin{equation}\label{ateos}
	\frac{a(t)}{a_0} = \left( \frac{9 \Omega_{\text{DM,0}}}{4} \right)^{1/3} \left[ (H_0  t)^{2/3} + \sqrt{w_0} \frac{2\pi}{9\sqrt{\Omega_{\text{DM,0}}} (H_0  t)^{1/3}} - w_0 \frac{24 + \pi^2}{81 \Omega_{\text{DM,0}} (H_0  t)^{4/3}}
	\right] + O(w_0^{3/2}).
\end{equation}
This is the desired equation expressing the evolution of the scale factor in the small pressure limit during matter domination for BEC dark matter.  We note that upon setting $w_0=0$ we immediately obtain the SCDM result: $a_{\text{SCDM}} \propto t^{2/3}$.  The additional terms are modifications taking into account the nonzero pressure of BEC dark matter.  

\begin{figure}
	\centering
		\includegraphics[width=4 in]{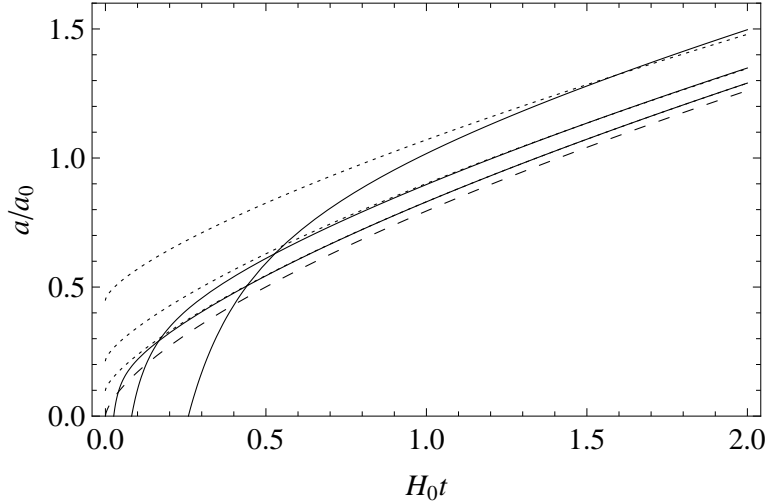}
	\caption{The evolution of the scale factor is plotted during an epoch of matter domination.  The dashed line corresponds to the SCDM result.  The solid lines are our approximate, analytical results (\ref{ateos}) for BEC dark matter.  From top to bottom they correspond to $w_0 = 0.1$, $0.01$ and $0.001$.  The dotted lines are the exact results (\ref{ateosexact}) for the same values of $w_0$.}
\label{fig:scalefactor}
\end{figure}

Using the WMAP result $\Omega_{\text{DM},0}=0.228$ \cite{wmap}, we have plotted the scale factor in figure \ref{fig:scalefactor}.
The dashed line on the bottom is the SCDM solution.  The solid lines are our approximate, analytical solutions (\ref{ateos}) for various $w_0$ (see caption) and the dotted lines the exact solutions (\ref{ateosexact}) for the same $w_0$.  As can be seen, as $H_0 t$ gets smaller the analytical solutions become less accurate.  The reason for this can be traced to our assumption that $y$ in (\ref{yeq}) becomes large for small $w_0$.  As $a/a_0$ decreases, this assumption becomes less valid.  In (\ref{ateos}) this manifests itself as the magnitude of the $w_0$ terms increasing for smaller $H_0 t$, making the truncated expansion in (\ref{ateos}) less valid.  Since present-day has been defined as $a/a_0 = 1$, we can see that BEC dark matter leads to a larger scale factor than for SCDM.


\section{Inhomogeneities}
\label{inhomogeneities}

In section \ref{sec:homogcos} we considered the homogeneous, isotropic Universe.  In this section we perturb around this Universe with the goal of determining the evolution of the perturbations, or inhomogeneities, to the energy density of BEC dark matter.  We begin in the following subsection with inhomogeneities in Newtonian gravity.  For nonrelativistic matter and subhorizon perturbations, Newtonian gravity is (nearly) sufficient for determining the leading order solutions \cite{muk}.  In section \ref{inhomgGR} we solve for the evolution of inhomogeneities in General Relativity.  While the use of the complete gravitational theory of General Relativity has the benefit of verifying our Newtonian solutions, it also introduces relativistic corrections and allows for the consideration of superhorizon perturbations, which Newtonian gravity is incapable of describing.  While we could present only the General Relativistic results, we find the Newtonian analysis more transparent and some discussions in section \ref{inhomgNewton} are necessary for justifying a straight forward application of General Relativity in section \ref{inhomgGR}. 


\subsection{Inhomogeneities in Newtonian Gravity}
\label{inhomgNewton}

In the absence of perturbations, the Universe is well known to be homogeneous and isotropic on the distance scales of interest \cite{muk}.  This means the energy density of matter does not vary over space and the velocity of matter is due only to expansion and obeys the Hubble law:
\begin{equation}
	\rho_0=\rho_0(t), \qquad \vect{v}_0 = \vect{v}_0(t,\vect{x}) = H(t)\vect{x},
\end{equation}
where the subscripted 0's indicate that these are background, unperturbed quantities.  Before perturbing them, there are two important equations we can derive.  Plugging these quantities into the continuity equation (\ref{eqcontin}) and the divergence of the Euler equation (\ref{eqsb}) combined with the Poisson equation (\ref{poiseq}) we find
\begin{equation} \label{nrFriedmann}
	\dot{\rho}_0 = -3H\rho_0, \qquad \dot{H}+H^2 = -\frac{4\pi G}{3c^2}\rho_0,
\end{equation}
which are the Friedmann equations and, as promised in section \ref{sec:eos}, we have derived them directly from (\ref{eqs}).  Note that compared to (\ref{Friedmann}), the Friedmann equations here are missing the pressure term.  This is because these are the nonrelativistic Friedmann equations, derived within Newtonian gravity.  If we were to write our equations in terms of the mass density instead of the energy density, then, as can be seen from (\ref{Friedmann}), only the pressure would contain factors of $c$, showing that it is a relativistic correction.  An alternative form for these equations will soon be of use.  Below we will introduce the scale factor, $a(t)$, which will be related to the Hubble parameter by $H = \dot{a}/a$, where a dot denotes a time derivative.  In terms of the scale factor the two equations in (\ref{nrFriedmann}) can be written
\begin{equation} \label{nrFriedmann2}
	\dot{a}^2 = \frac{8\pi G a^2}{3c^2} \rho_0, \qquad \ddot{a} = -\frac{4\pi Ga}{3c^2}\rho_0.
\end{equation}

We now perturb the above quantities around their background values:
\begin{subequations}
\begin{align}
	\rho(t,\vect{x}) &= \rho_0(t) + \delta\rho(t,\vect{x}) \\
	\vect{v}(t,\vect{x}) &= \vect{v}_0(t,\vect{x}) + \delta\vect{v}(t,\vect{x}) = H(t)\vect{x}  + \delta\vect{v}(t,\vect{x})\\
	V_G(t,\vect{x}) &= V_{G0}(t,\vect{x}) + \delta V_{G0}(t,\vect{x}).
\end{align}
\end{subequations}
In terms of these perturbed quantities, (\ref{eqs}) becomes
\begin{subequations} \label{eqs2}
\begin{align}
	(\delta\dot\rho)_{{x}} &= -\rho_0 \nabla_{{x}} \cdot \delta\vect{v} - \nabla_{{x}} \cdot (\delta\rho \, \vect{v}_0) \\
	-\delta\vect{\dot{v}} &= -\frac{\hbar^2}{2m}\frac{1}{\rho_0} \nabla_{{x}} (\nabla_{{x}}^2 \delta\rho) + \nabla_{{x}} (\vect{v}_0\cdot\delta\vect{v}) + \frac{2c^2w}{\rho_0}\nabla_{{x}}\delta\rho + \nabla_{{x}}\delta V_G \\
	\nabla_{{x}}^2 \delta V_G &= \frac{4\pi G}{c^2} \delta\rho,
\end{align}
\end{subequations}
where we've used the unperturbed version of (\ref{eqs}) to cancel terms and $w$ was defined in (\ref{eos}).  Soon we will Fourier transform these equations to facilitate solving them.  However, in their present form they will mix Fourier modes.  To avoid this we move to comoving coordinates, $\vect{q}$, given by $\vect{x} = a(t) \vect{q}$, which requires transforming derivatives as \cite{muk}
\begin{equation}
	\left(\frac{\partial}{\partial t} \right)_{{x}} = \left(\frac{\partial}{\partial t} \right)_{{q}} - \vect{v}_0\cdot \nabla_{{x}}, \qquad \nabla_{{x}} = \frac{1}{a}\nabla_{{q}}.
\end{equation}
Then (\ref{eqs2}) becomes
\begin{subequations} \label{eqscm}
\begin{align}
	\dot\delta &= -\frac{1}{a}\nabla\cdot \delta\vect{v} \label{concm}\\
	-\delta\dot{\vect{v}} &= -\frac{\hbar^2}{2m}\frac{1}{a^3} \nabla(\nabla^2\delta) + H\delta\vect{v} + \frac{2c^2w}{ a} \nabla\delta + \frac{1}{a}\nabla\delta V_G \label{eulercm}\\
	\nabla^2\delta V_G &= \frac{4\pi G}{c^2} a^2\rho_0 \delta, \label{poiscm}
\end{align}
\end{subequations}
where 
\begin{equation}
	\delta \equiv \frac{\delta \rho}{\rho_0}
\end{equation}
is the density contrast. In (\ref{eqscm}) we refrained from writing subscripted $q$'s and, for the Euler equation (\ref{eulercm}), used the fact that the velocity of a BEC is irrotational, as follows from (\ref{rhoveldef}).  These equations may be combined by taking the divergence of the Euler equation (\ref{eulercm}) and then subbing into it the continuity equation (\ref{concm}) and the Poisson equation (\ref{poiscm}), giving
\begin{equation} \label{eqNewton}
	\ddot\delta = -\frac{\hbar^2}{2m} \frac{1}{a^4} \nabla^4 \delta - 2H\dot\delta + \frac{2wc^2}{a^2}\nabla^2\delta + \frac{4\pi G}{c^2}\rho_0\delta.
\end{equation}
This equation describes the evolution of the density contrast, $\delta$, in an expanding universe.  Our interest is to solve this equation and look for growing modes representing the growth of inhomogeneities.  Such inhomogeneities will eventually become nonlinear ($\delta > 1$) leading to galaxy formation.  We focus here on the linear regime ($\delta < 1$), where a perturbative analysis is accurate, paying particular attention to the rate at which inhomogeneities in BEC dark matter grow compared to SCDM.

To solve (\ref{eqNewton}) we begin by Fourier transforming the density contrast,
\begin{equation}
	\delta(t,\vect{x})=\int\frac{d^3 k}{(2\pi)^{2/3}} \delta_{k}(t) e^{i\vect{k}\cdot\vect{x}},
\end{equation}
so that (\ref{eqNewton}) becomes
\begin{equation} \label{eqNewtonF}
\ddot\delta_{k}+2H\dot\delta_{k} + \left( \frac{\hbar^2 }{2m} \frac{k^4}{a^4} + 2wc^2\frac{k^2}{a^2} - \frac{4\pi G}{c^2}\rho_0\right)\delta_{k} = 0.
\end{equation}
At present, to solve (\ref{eqNewtonF}) one needs to specify the exact time dependence of the scale factor $a(t)$.  To avoid this, we can transform the independent variable from cosmic time, $t$, to the scale factor, $a$, with the help of (\ref{nrFriedmann2}), to obtain \cite{chavanis}
\begin{equation} \label{eqNewtonFa}
	\frac{d^2 \delta_k}{da^2} + \frac{3}{2a}\frac{d \delta_k}{da} + \frac{3}{2a^2}\left(\frac{\hbar^2 k^4}{8\pi G m^2 a^4 \rho_0} + \frac{wc^4 k^2}{2\pi G a^2 \rho_0} - 1 \right) \delta_k = 0,
\end{equation}
where now $\delta_k(a)$, $\rho_0(a)$ and $w(a)$ are functions of the scale factor.

For large enough $k$, the solutions are oscillating sound waves.  From (\ref{eqNewtonF}) we can see that this corresponds to the gravity term being negligible and pressure dominating.  For small  enough $k$ the gravity term dominates and the solutions are growing or decaying.  Since we are interested in growing solutions, our focus is then on smaller modes.  Ideally we would solve (\ref{eqNewtonFa}) exactly.  Unfortunately, it does not appear possible to find exact solutions \cite{chavanis}.  The problem is due to the $k^4$ term, which comes from the quantum pressure term in (\ref{eqsb}).  At the end of section \ref{sec:hydro} we mentioned that the quantum pressure term often causes difficulty for finding analytical solutions and that neglecting it is known as the Thomas-Fermi approximation.  We will now employ this approximation.  Note that making the Thomas-Fermi approximation leads to the standard \textit{classical} hydrodynamic equations \cite{muk}.  We will make heavy use of this fact in the next subsection.

\begin{figure}
	\centering
		\includegraphics[width = 6 in]{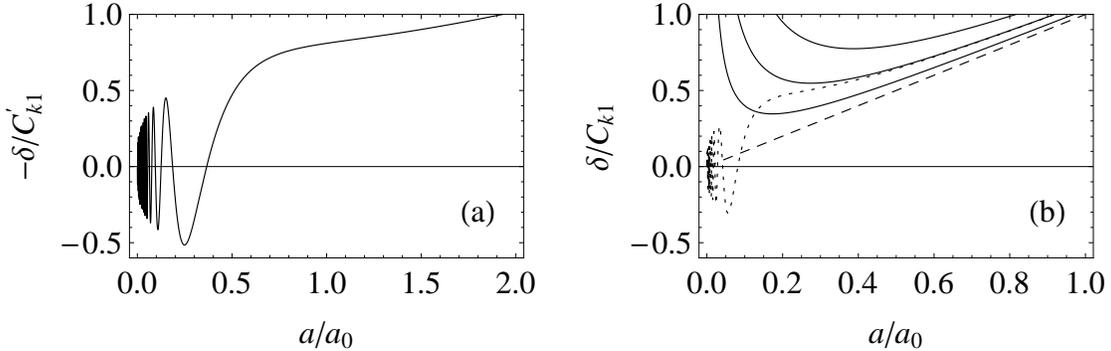}
	\caption{In (a) the exact solution (\ref{chavanisol}) is shown for $k=k_{J,0}$.  In (b) the dashed line is the $w_0=0$ SCDM solution, the solid lines are our approximate, analytical solutions (\ref{Newtonsola}) for, from top to bottom, $k^2/k_{J,0}^2=1/10$, $1/20$ and $1/50$.  The dotted line is the exact solution (\ref{chavanisol}) for $k^2/k_{J,0}^2=1/20$.}
\label{fig:newton}
\end{figure}

Making the Thomas-Fermi approximation, (\ref{eqNewtonFa}) can be written
\begin{equation} \label{eqNewtonF2}
	\frac{d^2 \delta_k}{da^2} + \frac{3}{2a}\frac{d \delta_k}{da} + \frac{3}{2a^2}\left[ 
	\left(\frac{a_0}{a}\right)^2 \left(\frac{k}{k_{J,0}}\right)^2 - 1	\right] \delta_k = 0,
\end{equation}
where $a_0$ is the present-day value of the scale factor and $k_J = a (2\pi G \rho_0 / wc^4)^{1/2}$ is the Jeans mode, with present day value
\begin{equation}
	k_{J,0} = a_0\sqrt{\frac{2\pi G \rho_0}{wc^4}} 
	= a_0 \left(\frac{3 \Omega_{\text{DM,0}} H_0^2}{4 c^2 w_0} \right)^{1/2}.
\end{equation}
The Jeans mode gives the exact point of crossover from oscillating solutions to growing and decaying solutions.  Oscillating solutions occur for modes $k > k_J$ while growing and decaying solutions occur for modes $k < k_J$.  Since $k_{J,0}$ is independent of $a$, all dependence on the scale factor (other than in $\delta_k$) has been explicitly written in (\ref{eqNewtonF2}).  The solution is \cite{chavanis}
\begin{equation} \label{chavanisol}
	\delta_k(a) = C'_{k1} \left(\frac{a_0}{a}\right)^{1/4}  J_{-5/4}\left[ \sqrt{\frac{3}{2}} \left(\frac{a_0}{a}\right) \frac{k}{k_{J,0}}\right]
	 + C'_{k2} \left(\frac{a_0}{a}\right)^{1/4} J_{5/4} \left[ \sqrt{\frac{3}{2}} \left(\frac{a_0}{a}\right) \frac{k}{k_{J,0}}\right],
\end{equation}
where $C'_{k1}$ and $C'_{k2}$ are arbitrary constants and $J_{\pm 5/4}$ are Bessel functions of the first kind.  In figure \ref{fig:newton}(a)
we have dropped the decaying solution ($C'_{k2}=0$) and plotted the growing solution for $k=k_{J,0}$.  For $a<a_0$, the solution is oscillating, since pressure is dominating over gravity, as can be seen from (\ref{eqNewtonF2}).  For $a>a_0$, gravity begins to dominate and the growing solution emerges.

As mentioned previously, the direction we take in this paper is to find analytical solutions in the small pressure limit that, upon setting $w_0 = 0$, reproduce SCDM results.  Such a solution can be obtained by expanding (\ref{chavanisol}), but we must be careful to expand around a small quantity.  Since growing solutions occur for $k<k_J$, we will focus on modes that satisfy $k \ll k_{J,0}$ and $k/k_{J,0}$ will be our small quantity.  For these modes, we find
\begin{equation} \label{Newtonsola}
	\delta_k(a) = C_{k1}\left[ \left(\frac{a}{a_0}\right) + w_0 \frac{3 k^2}{2\tilde{k}^2} \left(\frac{a}{a_0}\right)^{-1}  \right] + 
	C_{k2} \left[ \left(\frac{a}{a_0} \right)^{-3/2} - w_0 \frac{k^2}{6 \tilde{k}} \left(\frac{a}{a_0}\right)^{-7/2}
	\right]
	+ O(w_0^2),
\end{equation}
where $\tilde{k}^2 \equiv w_0 k^2_{J,0}$ (and is independent of $w_0$), $C_{k1}$ is an arbitrary constant proportional to $C'_{k1}$ and likewise for $C_{k2}$ and $C'_{k2}$.  One can see clearly that for $w_0 = 0$ this equation reproduces the SCDM solution: $\delta_{k,\text{SCDM}} = C''_{k1}a + C''_{k2}a^{-3/2}$ \cite{kolb}.

Dropping the decaying solution ($C_{k2}=0$) we have plotted the growing solution in figure \ref{fig:newton}(b).  The dashed line is the $w_0=0$ SCDM solution, the dotted line is the exact solution (\ref{chavanisol}) and the solid lines are our approximate, analytical solutions (\ref{Newtonsola}) for various values of $k^2/k_{J,0}^2$ (see caption).  Here again we find that the approximate solutions become less valid as $a/a_0$ decreases.  The reason is the same as before: as $a/a_0$ decreases, the $w_0$ term in (\ref{Newtonsola}) increases, lessening the validity of the truncated expansion.  We also find that inhomogeneities are larger with BEC dark matter.  In a universe with BEC dark matter, then, galaxy formation is expected to happen sooner than in the $\Lambda$CDM universe \cite{harko,chavanis}.

To write the solution (\ref{Newtonsola}) in terms of cosmic time, $t$, requires knowledge of how the scale factor evolves with time.  Since the evolution of the scale factor changes as different components dominate the total energy density, we must further specify a particular epoch of the Universe during which we wish to determine $a(t)$.  In (\ref{ateos}) we found $a(t)$ during matter domination when BEC dark matter dominates the total energy density.  It is customary to drop the purely decaying $C_{k2}$ solution so that, upon subbing (\ref{ateos}) into (\ref{Newtonsola}), we have
\begin{equation}\label{Newtonsolt}
\begin{split}
	\delta_{k}(t) =  C_{k1} & \left(\frac{9\Omega_{\text{DM},0}}{4}\right)^{1/3}
	 \Biggl[ (H_0t)^{2/3} + \sqrt{w_0} \frac{2\pi}{9\sqrt{\Omega_{\text{DM},0}}}(H_0 t)^{-1/3} \\
	 &+ w_0 \frac{k^2}{\tilde{k}^2} \left( \frac{2}{3 \Omega_{\text{DM,0}}^2}\right)^{1/3} (H_0 t)^{-2/3}
	 -w_0 \frac{24+\pi^2}{81\Omega_{\text{DM,0}}} (H_0 t)^{-4/3}	 
	  \Biggr] + O\left(w_0^{3/2} \right),
\end{split}
\end{equation}
Again, we find the SCDM solution when setting $w_0 = 0$.  

In (\ref{Newtonsolt}) the leading modification to the SCDM solution is positive.  Thus, we see analytically that BEC dark matter leads to an increased rate for the growth of inhomogeneities compared to SCDM.  This result is in line with \cite{harko,chavanis}, however in those papers the authors determined their complete solutions numerically.  Here we have obtained analytical solutions for growing modes during matter domination.  It is well known that the growth of inhomogeneities occurs at an appreciable rate only during matter domination \cite{muk}, and thus we have focused on this epoch.

In this subsection we have made a nonrelativistic, Newtonian analysis for the evolution of inhomogeneities.  However, in our final equation (\ref{Newtonsolt}) we used the relativistic result (\ref{ateos}) which followed from the fully relativistic Friedmann equations (\ref{Friedmann}) and not the Newtonian Friedmann equations (\ref{nrFriedmann}).  Using the Newtonian Friedmann equations would remove the $\sqrt{w_0}$ term.  The importance of this term, then, begs the question of how important relativistic corrections are for the density fraction.  One possibility would be to make a post-Newtonian analysis, as was done in \cite{harko,chavanis}.  Post-Newtonian gravity \cite{mccrea,lima} includes the leading relativistic corrections from General Relativity, and not just those in the Friedmann equations.  We opt instead to make a fully relativistic analysis using General Relativity.


\subsection{Inhomogeneities in General Relativity}
\label{inhomgGR}

From this point forward we set $c=1$.  We consider only a flat universe and scalar perturbations and write the metric in conformal Newtonian gauge as
\begin{equation} \label{metric}
	ds^2 = a^2(\eta)\left[(1+2\Psi) d\eta^2 - (1-2\Phi) \delta_{ij} dx^i dx^j \right],
\end{equation}
where $\Psi$ and $\Phi$ are scalar perturbations, $\eta$ is conformal time related to cosmic time via $dt = a d\eta$ and the $x^i$ are comoving coordinates.  The Hubble parameter, in terms of conformal time, is given by ${\cal H} = a'/a$, where a prime will denote differentiation with respect to $\eta$, and the Friedmann equations are
\begin{equation} \label{GRFriedmann}
	{\cal H}^2 = \frac{8\pi G}{3}a^2 \rho_0, \qquad {\cal H}' = - \frac{4\pi G}{3}a^2 \left(\rho_0 + 3p_0 \right),
\end{equation}
which may also be written as
\begin{equation}
	\left(\frac{a'}{a}\right)^2 =  \frac{8\pi G}{3}a^2 \rho_0, \qquad \frac{a''}{a} = \frac{4\pi G}{3}a^2(\rho_0 - 3p_0).
\end{equation}

The Hubble length, or horizon, is given by the physical distance $H^{-1} = (a{\cal H})^{-1} \sim a \eta$.  A perturbation with comoving mode $k \sim 1/\lambda$ has a physical wavelength of roughly $a\lambda$.  Thus $k\eta<1$ corresponds to superhorizon modes with physical wavelengths longer than the Hubble horizon and $k\eta > 1$ corresponds to subhorizon modes with physical wavelengths shorter than the Hubble horizon.  In the previous subsection we considered only subhorizon modes, since this is all a Newtonian analysis can accommodate.  In this subsection we consider both subhorizon and superhorizon modes.

The metric (\ref{metric}) obeys the Einstein field equations
\begin{equation}
	G_\mu^\nu = 8\pi G T_\mu^\nu,
\end{equation}
where the Einstein tensor $G_\mu^\nu$ is a function of the metric and $T_\mu^\nu$ is the stress-energy tensor.  The perturbed Einstein field equations, $\delta G_\mu^\nu =  8\pi G \delta T_\mu^\nu$, are \cite{muk}
\begin{subequations}\label{peree}
\begin{align}
	4\pi G a^2 {\delta T}_0^0 &= \nabla^2\Phi - 3{\cal H}\left(\Psi' + {\cal H}\Phi\right) \label{perefea}\\
	4\pi G a^2 {\delta T}_i^0 &= \left(\Psi' + {\cal H}\Psi\right)_{,i}  \label{perefeb}\\
	4\pi G a^2 {\delta T}_j^i &=  \frac{1}{2}(\Phi-\Psi)_{,ij}  - \left[\Psi'' + {\cal H} (2\Psi' + \Phi') + (2{\cal H}' + {\cal H}^2)\Phi + \frac{1}{2}\nabla^2(\Phi-\Psi) \right] \delta_{ij}, \label{perefec}
\end{align}
\end{subequations} 
where we have used the standard notation that a Latin index refers to spatial components only and a comma denotes a partial derivative, e.g. $\Psi_{,i} = \partial \Psi/\partial x^i$.  In this subsection, as above, we restrict our attention to the epoch of matter domination, during which BEC dark matter is dominating the total energy density of the Universe.  Then $T_\mu^\nu$ is the stress-energy tensor for dark matter.  As explained in the previous subsection, we are able to drop the quantum pressure term in the Euler equation (\ref{eqsb}) by making the Thomas-Fermi approximation.  This approximation reduces the equations (\ref{eqs}) to the standard \textit{classical} hydrodynamic equations of a perfect fluid.  Such equations may be derived from the stress-energy tensor for a perfect fluid,
\begin{equation} \label{setensorpf}
	T_\mu^\nu = (\rho + p)u^\nu u_\mu - p\delta_\mu^\nu,
\end{equation}
where $u^\mu$ is the 4-velocity of the fluid.  Conservation of the stress-energy tensor and the nonrelativistic limit reproduces exactly the continuity equation (\ref{eqcontin}) and the Euler equation (\ref{eqsb}) if one drops the quantum pressure term and makes the identification $p=V_0\rho^2/2m^2 c^4$, which is identical to the equation of state (\ref{BECeos}), which was derived through other means.  Perturbations to the perfect fluid stress-energy tensor are given by \cite{muk}
\begin{equation} \label{setensorpfper}
	\delta T_0^0 = \delta \rho, \qquad \delta T_0^i = a(\rho_0 + p_0) \delta v^i, \qquad \delta T_i^j = -\delta p \delta_i^j.
\end{equation}
Since the perturbations to the stress-energy tensor are diagonal, (\ref{perefec}) tells us $\Phi=\Psi$.  In the following, then, we shall label all scalar perturbations with $\Phi$.

BEC dark matter satisfies the equation of state (\ref{eos}), so that $\delta p = 2w  \delta \rho$.  Combining (\ref{perefea}) and (\ref{perefeb}) we obtain
\begin{equation} \label{greq1}
	\Phi''_k + 3(1+2w){\cal H}\Phi'_k + [2{\cal H}' + {\cal H}^2 + 2w ( 3{\cal H}^2 + k^2)]\Phi_k = 0,
\end{equation}
and from (\ref{perefea}) alone
\begin{equation} \label{greq2}
	\delta_k = -2 \left[ \frac{1}{{\cal H}}\Phi_k' +\left(1 + \frac{k^2}{{3\cal H}^2}\right) \Phi_k \right],
\end{equation}
where we made use of the Friedmann equations, $\delta_k = \delta \rho_k / \rho_0$ is the density contrast and we have Fourier transformed both $\Phi$ and $\delta$.  If we are able to solve (\ref{greq1}) for $\Phi_k$, then we may place it into (\ref{greq2}) to obtain $\delta_k$.  Note that, in general, the $\delta_k$ in (\ref{greq2}) is the density contrast for the perturbation to the \textit{total} energy density.  Since we are restricting ourselves to the epoch of matter domination, the total energy density is the energy density of dark matter and the $\delta_k$ in (\ref{greq2}) is the same $\delta_k$ that we solved for in the previous subsection, the inhomogeneity to BEC dark matter during matter domination.

To solve (\ref{greq1}) we first transform the independent variable from conformal time, $\eta$, to the scale factor, $a$, yielding
\begin{equation} \label{phide0}
	\frac{\partial^2 \Phi_k}{\partial a^2}  + \frac{1}{4a}\left(14 + 30 w \right)\frac{ \partial \Phi_k}{\partial a} + \frac{3 w}{a^2} \left(1 + \frac{k^2}{4\pi G a^2 \rho_0} \right)\Phi_k = 0.
\end{equation}
Using the results in section \ref{sec:homogcos} we can exchange $w$ for its present-day value, $w_0$, and obtain
\begin{equation} \label{phide}
	\frac{\partial^2 \Phi_k}{\partial a^2}  + \frac{1}{4a}\left[14 + w_0 \frac{30}{(a/a_0)^3} \right]\frac{ \partial \Phi_k}{\partial a} + \frac{w_0}{a_0^2}\frac{3}{(a/a_0)^5} \left(1 + \frac{1}{2} \frac{a}{a_0} \frac{k^2}{\tilde{k}^2}\right)\Phi_k + O(w_0^2)= 0.
\end{equation}
We could have written this equation in its entirety, and not just through order $w_0$.  However, it does not appear possible to solve the complete equation analytically.  Instead we solve the equation in the small pressure limit.  This allows use to solve for the solution expanded around small $w_0$.  We find
\begin{equation} \label{grphisol}
\begin{split}
	\Phi_k = C'_{k1} &\left\{1 + w_0\left[  \frac{3 k^2}{2\tilde{k}^2}\left(\frac{a}{a_0}\right)^{-2} - 2\left(\frac{a}{a_0}\right)^{-3} \right]
\right\} \\
&+ C'_{k2} \left\{\left(\frac{a}{a_0}\right)^{-5/2} - w_0 \left[\frac{ k^2}{6\tilde{k}^2} \left(\frac{a}{a_0}\right)^{-9/2} - \frac{21}{22} \left(\frac{a}{a_0}\right)^{-11/2} \right]\right\} + O(w_0^2),
\end{split}
\end{equation}
where $C'_{k1}$ and $C'_{k2}$ are arbitrary constants.  This equation gives the evolution of the gravitational potential for any $a$.  During matter domination we may write it in terms of cosmic time by using (\ref{ateos}).  It is customary to drop the purely decaying $C_{k2}$ solution so that we have
\begin{equation} \label{grphisoleta}
	\Phi_k = C'_{k1} \left\{1 + w_0 \left(\frac{4}{9\Omega_{\text{DM},0}}\right)^{2/3} \left[
		\frac{3 k^2}{2\tilde{k}^2} (H_0 t)^{-4/3} - 
		2\left(\frac{4}{9 \Omega_{\text{DM},0}} \right)^{1/3} (H_0 t)^{-2}\right]\right\} + O(w_0^{3/2}).
\end{equation}

During radiation domination, which preceded matter domination, $\Phi_k$ is constant for superhorizon modes \cite{dodelson}.  Since the Hubble horizon expands faster than the physical wavelength of a perturbation, the wavelength of the perturbation eventually becomes subhorizon.  If this occurs during radiation domination, the perturbation decays quickly to zero \cite{dodelson}.  If instead the perturbation survives into matter domination, then it evolves according to (\ref{grphisoleta}).  We find immediately, upon setting $w_0=0$, that (\ref{grphisoleta}) reproduces the SCDM solution of a constant gravitational potential for both superhorizon and subhorizon modes.  BEC dark matter introduces $t$-dependent corrections, so that the gravitational potential is no longer constant.

With the solution (\ref{grphisol}) for the gravitational potential we can obtain the density contrast using (\ref{greq2}).  To do so we first transform the independent variable in (\ref{greq2}) from conformal time to the scale factor and then, analogously to how we arrived at (\ref{phide}), we use the results in section \ref{sec:homogcos} to find
\begin{equation}
	\delta_k = -2 \left\{ a \partial_a\Phi_k +\left[1 + \frac{k^2}{4\tilde{k}^2} \frac{a}{a_0}  \left(1-w_0 +w_0\left(\frac{a}{a_0}\right)^{-3} \right)\right] \Phi_k \right\}.
\end{equation}
Into this equation we plug (\ref{grphisol}) to obtain the desired result:
\begin{equation} \label{grsola}
\begin{split}
	\delta_k &= C_{k1}\left\{\frac{a}{a_0} +  w_0 \left[ \frac{3k^2}{2\tilde{k}^2} \left(\frac{a}{a_0}\right)^{-1} - 7\left(\frac{a}{a_0}\right)^{-2}
	\right]
	+ \frac{4\tilde{k}^2}{k^2} \left[1+w_0+4w_0\left(\frac{a}{a_0}\right)^{-3}
	\right]
	\right\} \\
	&\quad +C_{2k}\Biggl\{
	 \left(\frac{a}{a_0}\right)^{-3/2} - w_0 \left[\frac{k^2}{6\tilde{k}^2} \left(\frac{a}{a_0}\right)^{-7/2} - \frac{283}{66} \left(\frac{a}{a_0}\right)^{-9/2}		\right]
	 \\
	 &\qquad \qquad - \frac{6\tilde{k}^2}{k^2} \left(\frac{a}{a_0}\right)^{-5/2}\left[1+w_0 + w_0\frac{63}{22}\left(\frac{a}{a_0}\right)^{-3} \right]
	\Biggr\}	+ O(w_0^2),
\end{split}
\end{equation}
where $C_{k1}=-k^2 C_{k1}'/2\tilde{k}^2(1-w_0)$ and likewise for $C_{k2}$ and $C_{k2}'$.  This result may be compared to the Newtonian result (\ref{Newtonsola}).  The fully relativistic solution (\ref{grsola}) reproduces the Newtonian terms (\ref{Newtonsola}) exactly.  We may rewrite (\ref{grphisoleta}) in terms of cosmic time by using (\ref{ateos}).  It is customary to drop the purely decaying $C_{2k}$ solution, so that we have
{\allowdisplaybreaks
\begin{equation} \label{grsoleta}
\begin{split}
	\delta_k 	&=C_{k1} \left(\frac{9 \Omega_{\text{DM},0}}{4} \right)^{1/3} \Biggl\{ (H_0 t)^{2/3} + \sqrt{w_0} \frac{2\pi}{9\sqrt{\Omega_{\text{DM},0}}} (H_0 t)^{-1/3} \\
		&\qquad + w_0 \left[ \frac{k^2}{\tilde{k}^2} \left(\frac{2}{3\Omega^2_{\text{DM},0}} \right)^{1/3} (H_0 t)^{-2/3} - \frac{24 + \pi^2}{81\Omega_{\text{DM,0}}} (H_0 t)^{-4/3}
		\right]\\
	&\qquad+ \frac{4\tilde{k}^2}{k^2} \left(\frac{4}{9\Omega_{\text{DM},0}}\right)^{1/3} \left[1+w_0 + w_0 \frac{16}{9\Omega_{\text{DM},0}} (H_0 t)^{-2}\right]
	\Biggr\}.
\end{split} 
\end{equation}
}%

During radiation domination, $\delta_k$ for SCDM is constant for superhorizon modes and grows at most logarithmically for subhorizon modes \cite{dodelson}.  The growth of inhomogeneities is largest during matter domination, during which $\delta_k$ for BEC dark matter evolves according to (\ref{grsoleta}) in the small pressure limit.  For $w_0=0$, (\ref{grsoleta}) reproduces the SCDM solution.  The dominant additional terms are positive and thus the growth rate of inhomogeneities increases for BEC dark matter compared to SCDM.  For subhorizon modes, $k\eta \sim kt^{1/3} \gg 1$ and we can drop the $k^{-2}$ terms.  This case was already  discussed in the previous subsection.  For superhorizon modes, $k\eta \sim kt^{1/3} \ll 1$ and we can drop all but the $k^{-2}$ terms:
\begin{equation}
	\delta_k = - 2C'_{k1} \left[1+2w_0+w_0 \frac{16}{9\Omega_{\text{DM},0}} (H_0 t)^{-2} \right] \quad \text{(superhorizon)}.
\end{equation}
From this solution we can see that during matter domination superhorizon inhomogeneities for SCDM are constant, while for BEC dark matter there is growth.


\section{Conclusion}
\label{conclusion}

In this work we studied the resulting cosmology when dark matter is a scalar field that has undergone Bose-Einstein condensation.  Such a model of dark matter has been shown to be in better agreement with the density profiles of galactic halos than standard cold dark matter in the $\Lambda$CDM model. We focused on the growth of inhomogeneities, i.e. perturbations to the dark matter energy density.  Since, as is well known, such perturbations only grow appreciably during matter domination, we considered only this epoch.  

Such an analysis had been considered in earlier work \cite{harko,chavanis} within the context of Newtonian and post-Newtonian gravity.   Our analysis differed from these papers principally in two ways.  First, the direction we took was to derive simple, analytical formulas that clearly showcased the modifications BEC dark matter produces in the standard cold dark matter solutions.  Second, while we did make our analysis first within Newtonian gravity, we then used the complete theory of General Relativity, finding solutions for the scale factor, gravitational potential and density contrast.  We found analytically that each one of these quantities increases at a faster rate compared to when dark matter is in the form of standard cold matter, consistent with the numerical results in \cite{harko,chavanis}.
  
Our fully relativistic solutions for the gravitational potential in (\ref{grphisoleta}) and the density contrast in (\ref{grsola}) are valid for both subhorizon and superhorizon perturbations, the latter of which is beyond the reach of Newtonian gravity.  These fully relativistic solutions are also necessary  for studying anisotropies in the radiation spectrum.  While we did not include an analysis of anisotropies here, such an analysis we expect to lead to new and interesting physics and would rely on our results.


\section*{Acknowledgments}

H. Y. L. was supported in part by the US National Science Foundation and US Army Research Office.


\end{document}